\documentclass[useAMS,usenatbib,usegraphicx]{mn2e}
\def\aj{AJ}%
\def\araa{ARA\&A}%
\def\apj{ApJ}%
\def\apjs{ApJS}%
\def\apss{Ap\&SS}%
\def\aap{A\&A}%
\def\aaps{A\&AS}%
\def\mnras{MNRAS}%
\def\memras{MmRAS}%
\def\pasp{PASP}%
\def\procspie{Proc.\,SPIE}%

\title[HIP\,45314 A+B]{The low mass companion of HIP\,45314 (HR\,3672)\thanks{Based on data products from observations made with ESO Telescopes at the La Silla Paranal Observatory under programme IDs 074.D-0180(A) and 080.D-0348(A).} } 
\author[Adam, C. et al.]{C. Adam$^{1}$\thanks{E-mail:
christian.adam.1@uni-jena.de}, R. Neuh\"auser$^{1}$\thanks{E-mail:
rne@astro.uni-jena.de}, M. Mugrauer$^{1}$, J. G. Schmidt$^{1}$, T. O. B. Schmidt$^{1}$
\\
$^{1}$Astrophysikalisches Institut und Universit\"ats-Sternwarte, Schillerg\"asschen 2-3, D-07745 Jena, Germany}

\begin{document}
\date{}

\pagerange{\pageref{firstpage}--\pageref{lastpage}} \pubyear{2012}

\maketitle

\label{firstpage}

\begin{abstract}

We report the discovery of a very low-mass companion to HIP\,45314 (HR\,3672) located about $2.7\arcsec$ north-west of HR\,3672\,A. With four years of epoch difference between the two observations obtained with the Very Large Telescope we can reject by more than 4\,$\sigma$ that B would be a non-moving background object unrelated to A. HR\,3672\,A is a B\,7-8 main sequence star with an age of $\rmn{140\pm80\,Myr}$ and from the magnitude difference between HR\,3672\,A and B and the 2MASS magnitudes, we can estimate the magnitude of HR\,3672\,B ($\rmn{K_{s}=12.42\pm0.28\,mag}$) and then, for age and distance of the primary star, using models, its luminosity and mass ($\rmn{0.2-0.5\,M_{\sun}}$). We present a reanalysis of the probable membership of HR\,3672\,A to the Platais 9 cluster, and introduce a new method for astrometric calibration of data without dedicated calibration images. In the deepest available image (co-add of all epochs) no additional companion candidates within $\rmn{13\arcsec}$ were detected down to $\rmn{0.03\,M_{\sun}}$.

\end{abstract}


\section{Introduction}
It is still not clear, whether massive stars form preferentially by accretion of material through a circumstellar disk or by coagulation of lower mass stars in a multiple system. Maybe, both ways of formation are possible, then it is still unknown, how often each of the two channels is chosen by nature. High and intermediate mass stars very often are multiple \citep{2007ARA&A..45..481Z} with a binary fraction of 70 to 90 \% \citep{2007A&A...475..875P,2002MNRAS.336..309O}. Yet, for many high and intermediate mass stars, no deep and sensitive multiplicity survey has been done. In particular, only few of these stars have been observed so far by AO imagers, which can find very faint, maybe sub-stellar, companions also at a separation range (roughly 0.1 to few arc sec, i.e. long orbital periods), which is not reachable for spectroscopic surveys (short periods).
This object is part of our study of multiplicity of B-type stars in European Southern Observatory (ESO) archival Adaptive Optics (AO) imaging data. With this project we can study the statistical aspects of multiplicity among intermediate mass stars, analyse their distribution with respect to mass and separation and obtain constraints on the underlying processes in their formation. Here we show first results of our project.\\
The star HIP\,45314 (also called HR\,3672, distance $161.3\pm6.7\,\rmn{pc}$ at $\rmn{\alpha}(J2000)=09^{\rmn{h}} 14^{\rmn{m}} 08\fs2$ and $\rmn{\delta}\,(J2000)=-44\degr 08\arcmin 45\arcsec$ according to \cite{2007A&A...474..653V}, $\mathrm{V=5.839\,mag}$ according to Simbad) is located in the Vela constellation. \cite{1998AJ....116.2423P} identified HR\,3672 as a possible member of the open cluster HIP\,45189 (no probability of membership given) with an age of $\mathrm{\sim100\,Myr}$ (for details, see Secion\,\ref{mass_age_A}). The spectral type of HR\,3672 is B6IV according to Simbad, but as reported in the literature by different authors \citep[and summarized e.g. by][]{2010yCat....102023S}, mainly yielded spectral types between B4V and B7V (see Sec.\,\ref{spec_class} and Sec.\,\ref{mass_age_A}). The star is also suspected to be variable with an amplitude in V of $\mathrm{0.05\,mag}$ \citep{1981NVS...C......0K}.\\
We also use this object exemplary to introduce an alternative method for the astrometric calibration (self-calibration of pixel scale and detector orientation) of data by using the science images instead of extra recorded calibration binaries or clusters.\\
The order of this paper is as follows. In Sec.\,\ref{obs} the observations and data reduction are presented. A detailed description of the self-calibration for astrometry is given in Sec.\,\ref{calib}. Furthermore we show the result of the common proper-motion pair analysis in Sec.\,\ref{astrometry} and the age and mass estimation for HR\,3672\,A and B in Sec.\,\ref{mass_age_A} and Sec.\,\ref{mass_B}, respectively. The limits on detection of further companions are explained in Sec.\,\ref{det_limit} and the conclusions are collected in Sec.\,\ref{conclusion}. 


\section{Observations and data reduction}
\label{obs}
The data presented here are part of an archival search for unknown companions of B-type stars within 1\,kpc. We reduced all publicly available data of B-type stars taken with the Adaptive Optics camera NACO (for Nasmyth Adaptive optics system with the COude near-infrared imager and spectrograph, \citealt{2003SPIE.4839..140R, 2003SPIE.4841..944L}) located in a VLT/UT4 (Yepun) Nasmyth focus.

\begin{table}
\caption{Observations log}
\label{tab1}
\begin{tabular}{@{}lccccc}
\hline
Epoch & Optics \& Filter & {\em DIT} [s]& {\em NDIT} & {\em N}& {\em FW}\\
\hline
2004-12-16&$\mathrm{S27/K_{S}+ND_{Short}}$&0.3454&149&9&90\\
2008-02-15&$\mathrm{S27/K_{S}+ND_{Short}}$&1.0&50&9&101\\
\hline
\end{tabular}

\medskip
{\bf Note.} With detector integration time ({\em DIT}), the number of detector integrations per frame ({\em NDIT}), the total number of frames ({\em N}), and the full width ({\em FW}) at half maximum in milli arc seconds are given.
\end{table}

\begin{figure}
 \includegraphics[angle=0,width=84mm]{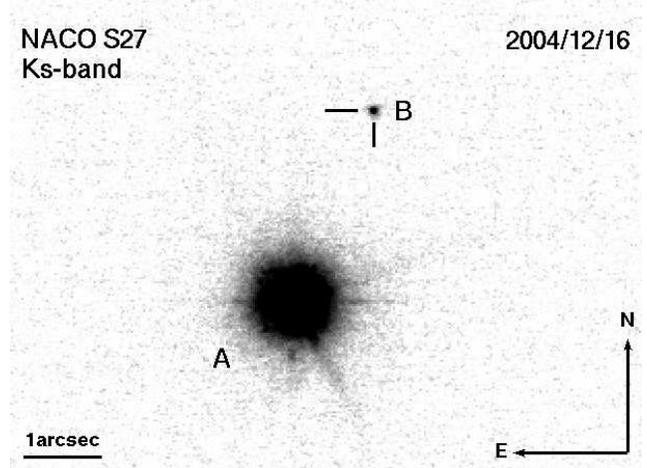}
  \caption{VLT-NACO S27 image of HR\,3672 and its faint companion candidate ($\Delta K_{\mathrm{s}} = 6.4\,\rmn{mag}$) taken on 2004-12-16 in Ks-band. The total integration time is $7.7\,\rmn{min}$. Since the companion candidate is found to be co-moving (Sec.\,\ref{astrometry}), we can call it B.}
  \label{NACO_img}
\end{figure}

The data-set of HR\,3672 consists of two epochs, one observed in 2004\footnote{Public data from the ESO data archive, taken in ESO program 074.D-0180(A)} with a total integration time of $7.7\,\rmn{min}$, and one from 2008\footnote{Public data from the ESO data archive, taken in ESO program 080.D-0348(A)} with a total integration time of $7.5\,\rmn{min}$ (for details see Table \ref{tab1}). In addition calibration data, i.e. darks with same exposure time as science frames, and lampflats with same filter as science frames, were taken from archive. Master-dark and flat field images for each epoch were created with \verb"esorex"\footnote{http://www.eso.org/sci/software/cpl/esorex}, which is part of ESO's Common Pipeline Library\footnote{http://www.eso.org/sci/software/cpl} (CPL). After removal of bad pixel with \verb"IDL/sigma_filter" \citep{1993ASPC...52..246L} from each science image, darks and flats were applied using ESO \verb"ECLIPSE/jitter"\footnote{ESO C Library for an Image Processing Software Environment.}. Finally we co-added all images with \verb"esorex", using the provided shift+add procedure. The fully reduced image, taken in December 2004, is shown in Figure\,\ref{NACO_img}.

To obtain the position of HR\,3672\,A and its companion candidate (henceforth, B) for all epochs we used \verb"MIDAS" center/gauss \citep{1992ASPC...25..115W} on the reduced images. Neither A nor B are saturated.\\
In order to determine the spectral type of HR\,3672 we also used a well exposed, so far unpublished spectrum of the star, which was obtained by the International Ultraviolet Explorer (IUE) on 1993 Feb 13th. The spectrum was taken with IUE's short-wavelength spectrograph (1150-1980\,\AA) and its prime camera (SWP) in the high resolution mode ($\Delta \lambda\sim 0.2$\,\AA, and an average dispersion of about 0.04\,\AA/pixel), using the large aperture (slot width $10\,''\times 20\,''$). The total integration time of the spectrum is 434.865\,s.


\section[calibration]{Astrometric calibration}
\label{calib}
Usually astrometric calibrations are obtained from binaries or clusters with known separations and position angles, which are observed with the same setup in the same night. In some cases neither calibration binaries nor clusters were observed for a data-set or it is very time-consuming to find adequate calibrations. In these cases it is still possible to achieve an acceptable astrometric calibration using a technique which we call self-calibration. A first limited usage was performed in \citet{2008IAUS..248..126S} and \citet{2010A&A...516A.112N}. In this section we now give a detailed description of the approach to this method using the example of HR\,3672.

The basic idea of the self-calibration is to obtain the pixel scale ($PS$) and the rotation angle ($\rmn{\theta}$) of the detector from the science images itself using the non-shifted and non-added reduced images. Two information's are needed for the calculation:

\begin{enumerate}
\renewcommand{\theenumi}{(\arabic{enumi})}
  \item position of objects on the detector $x_{\rmn{i}}$, $y_{\rmn{i}}$ in pixel-units
  \item $RA$ and $DEC$ coordinates of the telescope pointing, stored in the FITS-header keywords \emph{CRVAL1} and \emph{CRVAL2}
\end{enumerate}

For object detection we first used \verb"SExtractor" \citep{1996A&AS..117..393B} to get a raw position of the objects on the detector. We then passed the results to \verb"IDL/starfinder" \citep{2000SPIE.4007..879D} to obtain a more accurate object detection. From the resulting set of pixel positions $\{x_{\rmn{i}}$, $y_{\rmn{i}}\}_{\rmn{k}}$ ($\rmn{k}=1\,...\,\rmn{N}$, where $\rmn{N}$ is the total number of frames) we than calculate the shift of an object on the detector between each frame by
\[
  \Delta x \stackrel{\rmn{i}\neq \rmn{j}}{=} x_{\rmn{i}} - x_{\rmn{j}}, 
  \qquad \Delta y \stackrel{\rmn{i}\neq \rmn{j}}{=} y_{\rmn{i}} - y_{\rmn{j}}.
\]
The result is a new set of variables $\{\Delta x, \Delta y\}_{\rmn{m=1}}^{\rmn{n}}$ in which the number of sets $\rmn{n}$ is given by
\[
  \rmn{n} = \frac{\rmn{N} (\rmn{N} - 1)}{2}.
\]
Hence, for HR\,3672 ($\rmn{N}=9$ for both epochs as number of frames, i.e. of different telescope offsets) we get $\rmn{n}=36$ pairs of values, assuming that at least one object, namely the brightest, was detected in each frame. The set of coordinate values $\{\Delta RA, \Delta DEC \}_{\rmn{m=1}}^{\rmn{n}}$, which gives the shift of the telescope pointing, is calculated by
\[
  \Delta RA \stackrel{\rmn{i}\neq \rmn{j}}{=} -(RA_{\rmn{i}} - RA_{\rmn{j}}) \cos(\overline{DEC})
\]
and
\[
  \Delta DEC \stackrel{\rmn{i}\neq \rmn{j}}{=} DEC_{\rmn{i}} - DEC_{\rmn{j}},
\]
where $\overline{DEC}$ is the average of $DEC_{\rmn{i}}$ and $DEC_{\rmn{j}}$, and $\cos(\overline{DEC})$ is the cosine-correction between the 3D world coordinate system (WCS) and the 2D image coordinate system. From this we then can derive the separation $D_{\rmn{Det.}}$ on the detector and $D_{\rmn{WC}}$, the separation in world coordinates, by taking the root of the sum of squares. The pixel scale is then given by
\begin{equation}
  PS_{\rmn{m}} = \frac{D_{\rmn{WC,m}}}{D_{\rmn{Det.,m}}} \qquad \rmn{for}\, m = 1 ... \rmn{n}.
\end{equation}

\begin{figure}
 \includegraphics[width=54mm]{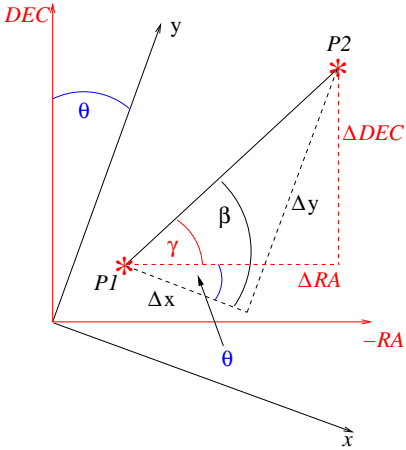}
\caption{Schematic diagram for the rotation angle of the detector $\rmn{\theta}$. The points $P1$ and $P2$ indicate the position of the object in two frames. The difference between the frames in detector coordinates is shown by $\Delta x$, $\Delta y$ and their corresponding angle $\beta$, as well as the difference in world coordinates ($\Delta RA$, $\Delta DEC$) with the angle $\gamma$. The rotation angle $\rmn{\theta}$ is calculated as difference of $\beta$ and $\gamma$; this angle gives the detector orientation.} 
\label{rot_ang_scheme_img}
\end{figure}

For the calculation of the detector rotation angle $\rmn{\theta}$ we use the fact that the amount of de-rotation is given by the difference of the rotated detector, here labelled as $\beta$, against the angle of the non-rotated world coordinate system, labelled as $\gamma$. Fig. \ref{rot_ang_scheme_img} shows a scheme of the de-rotation of the two coordinate systems. Hence, we get
\begin{equation}
  \rmn{\theta}_{\rmn{m}} = \beta_{\rmn{m}} - \gamma_{\rmn{m}}\qquad \rmn{for}\, m = 1 ... \rmn{n},
\end{equation}
where $\beta_{\rmn{m}}$ and $\gamma_{\rmn{m}}$ are given by
\[
 \beta_{\rmn{m}} = \arctan(\frac{\Delta y}{\Delta x}),
 \qquad   \gamma_{\rmn{m}} = \arctan(\frac{\Delta DEC}{\Delta RA}).
\]

\begin{figure}
 \includegraphics[width=84mm]{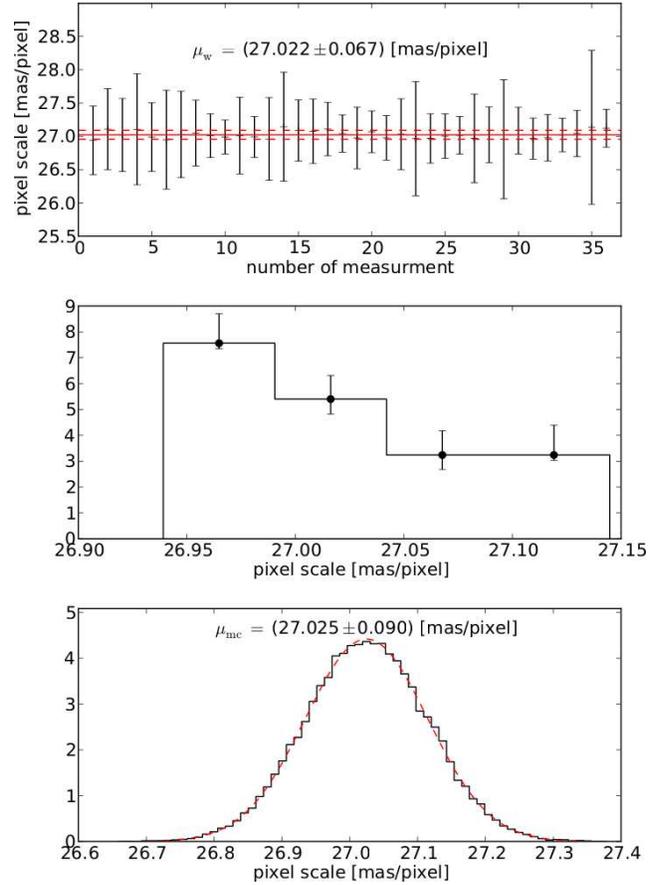}
\caption{We show the results from the self-calibration method. In the upper panel we show the pixel scales and the weighted mean of the values ($\mu_{\rmn{w}}$); the distribution of the data-set (middle panel), with bin-sizes calculated by Freedman-Diaconis choice \citep{fre81} and errors from deviation of re-sampled data-points with $15000$ iterations. The lower panel shows the histogram of the artificially increased sample with $50000$ iteration steps using Monte-Carlo. The best fit for normal distributed values is indicated by the (red) dashed line, with the mean pixel scale $\mu_{\rmn{mc}}= 27.025\pm0.090\,\rmn{mas/pixel}$.}
\label{ps_calib_2004_img}
\end{figure}

From this sample of data we excluded those values by $3\sigma$-clipping, whose deviations are $>3\,\sigma$ around the mean. In Fig.\,\ref{ps_calib_2004_img} (upper panel) we show the result from the pixel scale calculation for December 2004 as function of the measurement number. All errors are calculated by largest error-estimation. Since this method is based on relative distances, we use the tracking-error of the VLT/UT4 (Yepun) of $0.1\arcsec$\footnote{RMS of tracking-error for all VLT/UTs according to ESO} for error estimation. Furthermore we calculated the weighted mean of the data-set (full and dashed lines in Fig. \ref{ps_calib_2004_img} upper panel) and show the distribution of the data (middle panel). Because the number of data is a low-number statistic we artificially increased the sample with a Monte-Carlo simulation (for details of principal see, e.g. \cite{2003psa..book.....W}) with $50000$ iterations. We assumed that the underlying distribution is a Gaussian. To test the significance of the null hypothesis we used a Shapiro-Wilk test for normality \citep{Shap1965}. The results show that the null hypothesis can be accepted at a significance level of $>95\%$. With this assumption we generated normal distributed values including the errors from our data-set, which gave us the final result (mean and standard deviation) for the pixel scale $PS$ and the detector rotation angle $\rmn{\theta}$ shown in Tab. \ref{tab2}. 

\begin{table}
\caption{Results of self-calibration for HR\,3672}
\label{tab2}
\begin{tabular}{@{}lccc}
\hline
Epoch & Optics & $PS$ [mas/pixel] & $\rmn{\theta}$ [$\degr$] \\
\hline
2004-12-16 & S27 & 27.025$\pm$0.090 & 0.033$\pm$0.191 \\
2008-02-15 & S27 & 27.200$\pm$0.104 & 0.017$\pm$0.218 \\
\hline
\end{tabular}

\medskip
{\bf Note.} We list the pixel scale ($PS$) and the rotation angle ($\rmn{\theta}$) of the detector and their uncertainties for the used epochs.
\end{table}

\begin{table*}
\caption{Comparison of our new self-calibration for GQ Lup with normal astrometric calibration with a binary}
\label{tab2t}
\begin{tabular}{@{}lccccc}
\hline
GQ Lup & \multicolumn{2}{c}{\citep{2008A&A...484..281N}} & \multicolumn{2}{c}{this work, self-calibration}\\
Epoch & $PS$ [mas/pixel] & $\rmn{\theta}$ [$\degr$] & $PS$ [mas/pixel] & $\rmn{\theta}$ [$\degr$]\\
\hline
2005-05-27 & $13.240 \pm 0.050$ & $0.21 \pm 0.33$ & $13.221 \pm 0.059$ & $0.23 \pm 0.24$ \\
2005-08-08 & $13.250 \pm 0.052$ & $0.35 \pm 0.34$ & $13.279 \pm 0.056$ & $0.14 \pm 0.24$ \\
2006-02-22 & $13.238 \pm 0.053$ & $0.18 \pm 0.35$ & $13.205 \pm 0.063$ & $0.12 \pm 0.28$ \\
2006-05-20 & $13.233 \pm 0.055$ & $0.39 \pm 0.36$ & $13.172 \pm 0.049$ & $0.22 \pm 0.23$ \\
2006-07-16 & $13.236 \pm 0.055$ & $0.43 \pm 0.36$ & $13.241 \pm 0.059$ & $0.06 \pm 0.22$ \\
2007-02-19 & $13.240 \pm 0.059$ & $0.34 \pm 0.38$ & $13.304 \pm 0.128$ & $0.06 \pm 0.55$ \\
\hline
\end{tabular}

\end{table*}

To test our self-calibration technique we use six data-points published in \citet{2008A&A...484..281N} calibrated with the binary HIP\,73357, all with NACO S\,13. We re-reduced the science data and used our method to calculate pixel scale and detector orientation for each epoch. The results are shown in Tab. \ref{tab2t}. The comparison of the two methods shows that the results for the pixel scale and detector orientation are consistent within the $1\sigma$ errors. The errors on the detector orientation for our new self-calibration method are in most cases smaller than in \citet{2008A&A...484..281N}, but possible NACO field distortions could not be considered in our self-calibration. The fact that the pixel scale values on the GQ Lup data obtained from self-calibration scatter more than in \citet{2008A&A...484..281N} points to field distortions. Such distortions could be larger in S27 than in S13. But overall, our self-calibration should be consistent within the errors.\\
The presented method can be used in principal for all imaging techniques which use a dither method, but the accuracy of the results is strongly dependent on the tracking accuracy of the telescope, i.e. a ten times larger tracking-error for example would result in a ten times larger error in the rotation angle. Therefore this self-calibration should only be considered as rough estimation for data-sets without astrometric calibrators.


\section{Results}

\subsection{Astrometry}
\label{astrometry}

\begin{table}
\caption{Astrometry of HR\,3672\,A and its companion candidate}
\label{tab3}
\begin{tabular}{@{}lcccc}
\hline
Epoch & $\rho$ [$\arcsec$] & $\rho_{\mathrm{bg}}$ [$\arcsec$] & $s_{\mathrm{bg}}$ [$\sigma$] \\
\hline
2004-12-16 & 2.722$\pm$0.009 & 2.812$\pm$0.021 & 4.0\\
2008-02-15 & 2.742$\pm$0.021 & - & -\\
\hline
Epoch & PA [$\degr$] & PA$_{\mathrm{bg}}$ [$\degr$] & $s_{\mathrm{bg}}$ [$\sigma$]\\
\hline
2004-12-16 & 337.018$\pm$0.192 & 335.208$\pm$0.218 & 6.2\\
2008-02-15 & 336.331$\pm$0.218 & - & -\\
\hline
\end{tabular}

\medskip
{\bf Note.} Shown are the separation $\rho$ and the position angle PA (measured from North over East) of the companion candidate (B) with respect to HR\,3672\,A, the expected separation ($\rho_{\mathrm{bg}}$) and position angle (PA$_{\mathrm{bg}}$) if B would be a non-moving background object, and probability ($s_{\mathrm{bg}}$, in Gaussian $\rmn{\sigma}$) at which the hypothesis of a non-moving background object can be rejected.
\end{table}

With the results of the self-calibration (see above) we get the separation $\rho$ on the detector and the true, corrected position angle (PA) for HR\,3672\,A and its companion candidate, listed in Tab.\,\ref{tab3}. The companion candidate is located $\approx 2.73 \arcsec$ north-west of HR\,3672\,A (PA $\approx 336.7 \degr$). To check for common proper motion and orbital motion we use the astrometric data from Simbad: Proper motion $\mu_{\alpha}\cos(\delta) = -24.77\pm0.27\,\rmn{mas/yr}$ and $\mu_{\delta} = 13.10 \pm 0.20\,\rmn{mas/yr}$, distance $\mathrm{\pi =6.20\pm0.27\,mas}$ (or $161.3\pm6.7\,\rmn{pc}$), from \cite{2007A&A...474..653V}. Applying the Smith-Eichhorn \citep[ equation 21 therein]{1996MNRAS.281..211S} correction (to obtain the more reliable expectation value), we derive $\mathrm{\pi=6.26\pm0.27\,mas}$ or $\mathrm{159.7\pm6.6\,pc}$, respectively. The calculations presented here are done using this corrected parallax, but the results are almost identical for the Hipparcos parallax without Smith-Eichhorn correction.

\begin{figure}
 \includegraphics[width=8.4cm]{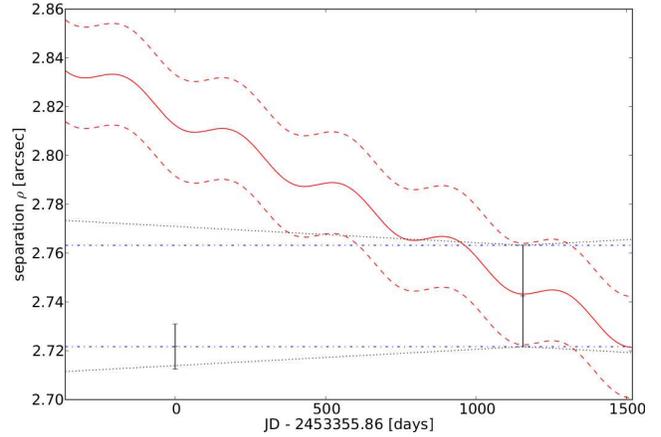}
  \caption{Plot of separation versus time-difference (in days) between observations for data listed in Tab. \ref{tab3}. Shown are the expected orbital motion for a circular edge-on orbit (dotted line); the background-hypothesis (full line), i.e. the expected change in separation assuming that the faint companion candidate is a non-moving background object, with errors including parallax and proper motion errors. The data point is inconsistent with the background hypothesis by a significance of $4\sigma$ and consistent with common proper motion (within dashed lines).}
  \label{sep_img}
\end{figure}

\begin{figure}
 \includegraphics[width=8.4cm]{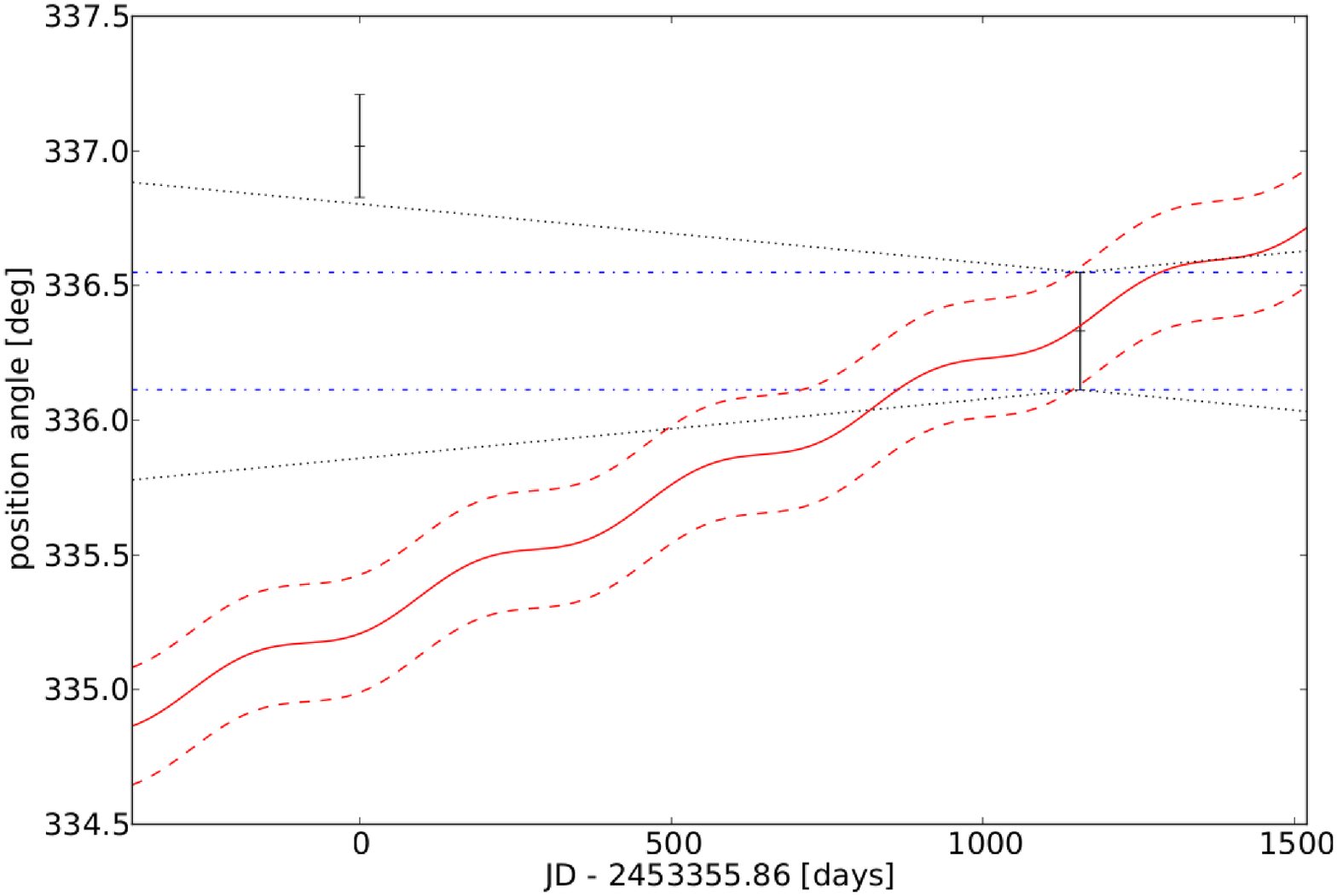}
  \caption{Plot of position angle versus time-difference (in days) between observations for data listed in Tab. \ref{tab3}. Shown are the expected orbital motion for a circular pole-on orbit (dotted line); the background-hypothesis (full line), i.e. the expected change in position angle assuming that the faint companion candidate is a non-moving background object, with errors including parallax and proper motion errors. The data point is inconsistent with the background hypothesis by a significance of $6\sigma$. The PA values are deviant by about 2$\sigma$, possibly due to orbital motion.}
  \label{pa_img}
\end{figure}

In the Figs. \ref{sep_img} and \ref{pa_img} we show the data from Tab.\,\ref{tab3}. We can reject the hypothesis that the companion candidate to HR\,3672\,A is a non-moving background object by $4\,\sigma\,\&\,6\,\sigma$ for the given epoch difference of $4\,\rmn{yr}$, and so we regard HR\,3672\,A and B as common proper motion pair. At the distance of HR\,3672 the projected physical separation of HR\,3672\,A+B would be $438\pm25\,\rmn{AU}$. We also show the expected maximum orbital motion for a circular orbit of HR\,3672\,B around A, being $\la 2.4\,\rmn{mas/yr}$ change in separation for an edge-on orbit (Fig.\,\ref{sep_img}) and $\la 0.08\,\rmn{\degr/yr}$ change in PA for a pole-on orbit (Fig. \ref{pa_img}). The coverage of the orbit (period $\rmn{4400\,yr}$ for circular orbit) from the two epochs is $<0.1\,\%$ so that we only can propose that HR\,3672\,B is orbiting around A. It is still possible that the two objects are uncorrelated, showing a similar 2D motion on sky. The PA values are deviant by about 2 $\sigma$, possibly due to orbital motion.

\subsection{Spectral Classification of HR\,3672\,A}
\label{spec_class}
To determine the mass of the companion, we need to know well the age and spectral type of the primary. Depending on literature, the given spectral type of HR\,3672 varies from B4V to B7V, see \citet{1969MNRAS.144...31B,1973MmRAS..77..199T,1973Obs....93...75W,1977A&AS...30...71C,1978mcts.book.....H}. In contrast to the other authors \citet{1969ApJ...157..313H} gives B6IV, i.e. lists HR\,3672 as non-main sequence star. This is not sufficient precise, so that we try to determine its spectral type with more precision.\\
In the spectral range, covered by the IUE observations (1150-1980\,\AA), the slope of the spectral continuum is expected to flatten from early to late type B stars, while the prominence of the Ly-$\alpha$-line (at 1216\,\AA) should strengthen towards later spectral types. Hence, these characteristics are well usable for spectral classification.

\begin{figure}
\resizebox{\hsize}{!}{\includegraphics{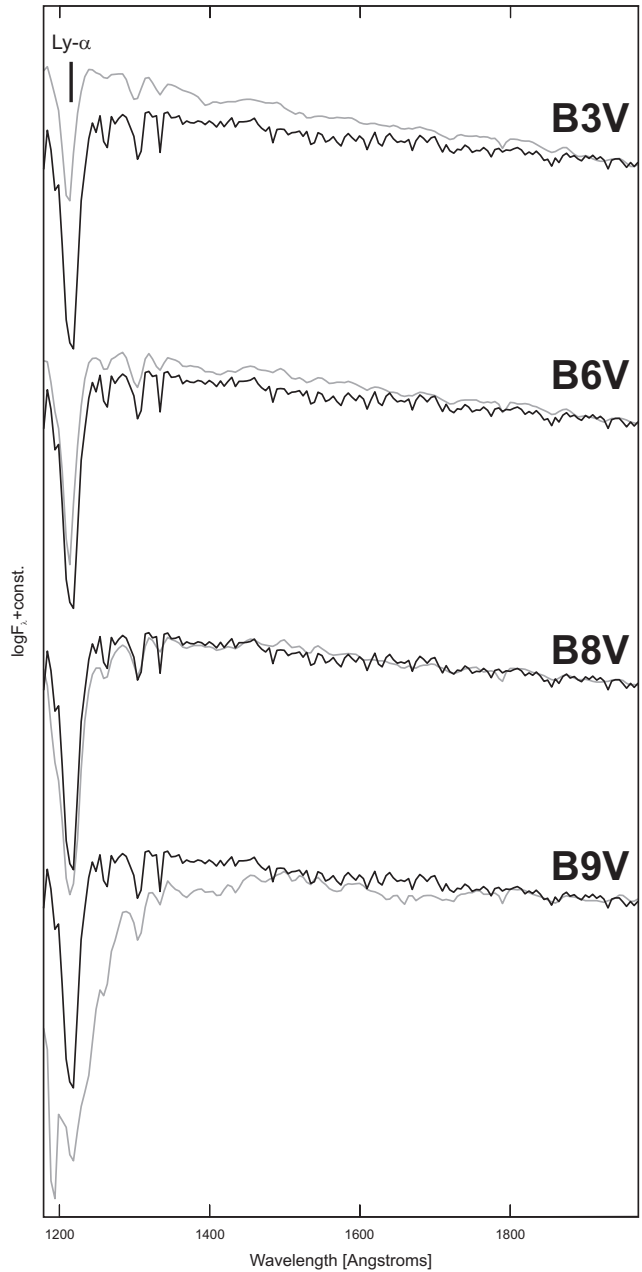}}
  \caption{The IUE spectrum of HR\,3672 (black) compared to template spectra (grey) of early to late type B stars from \citet{1998PASP..110..863P}. The spectrum of HR\,3672 is most consistent with a late type B dwarf with a spectral type between B7V to B8V.}
\label{specs}
\end{figure}

We compared the flux-calibrated IUE spectrum\footnote{obtained from the IUE archive http://sdc.laeff.inta.es/ines/} of HR\,3672 with template spectra from \cite{1998PASP..110..863P}. Thereby, we re-sampled the UV spectrum of HR\,3672 to match the resolution of the comparison spectra. The re-sampled IUE spectrum of HR\,3672 together with the spectral templates of early to late type B stars are shown in Fig.\,\ref{specs}.

We find that the slope of the spectrum of HR\,3672 is clearly flatter than the ones of the comparison spectra earlier than B6V, while it is steeper then the slope of the B9V template spectrum. Furthermore, the Ly-$\alpha$-line in the spectrum of HR\,3672 is more prominent than the Ly-$\alpha$-lines in the spectral templates earlier than B6V, while it is clearly weaker than the one of the B9V comparison spectrum. Hence, both the shape of the continuum, as well as the strength of the detected Ly-$\alpha$-line are consistent with a spectral type between B7V and B8V and we will exclude B6IV from our further investigations.
%

\subsection{Mass and Age of HR\,3672\,A}
\label{mass_age_A}

The B-type star HR\,3672 (HIP\,45314) has different published V-band and B-band magnitudes that are formally not consistent within their errors in each band. HR\,3672 might be a variable star \citep{2009yCat....102025S}, thus, we take the median value with standard deviation from the V-band magnitudes published in the last decade \citep[ see Tab.\,\ref{tabMag}]{2001KFNT...17..409K,2008AJ....136..735L,2006yCat.2168....0M,2001yCat.5109....0M,2006ApJ...638.1004A}, yielding $\mathrm{V=5.842\pm0.015\,mag}$. With the same procedure we obtain for the B-band magnitude $\mathrm{B=5.721\pm0.027\,mag}$. The JHK magnitudes are measured by 2MASS ($\mathrm{J=6.061\pm0.019\,mag}$, $\mathrm{H=6.154\pm0.033\,mag}$, $\mathrm{K_{s}=6.134\pm0.027\,mag}$), see \citet{2003yCat.2246....0C}. Using B and V magnitude we derive an intrinsic colour of $\rmn{(B-V)=-0.121\pm0.042\,mag}$, consistent with a B6-7 main-sequence star \citep{1995ApJS..101..117K}.

\begin{table}
\caption{Photometry data from Vizier for HR\,3672\,A}
\label{tabMag}
\begin{tabular}{@{}lcccccc}
 \hline
Bmag & Vmag & Ref. \\
\hline
- & 5.83$\pm$0.03 &  \cite{1976AAHam...9....1H} \\
5.731$\pm$0.003 & 5.842$\pm$0.003 &  \cite{2001KFNT...17..409K} \\
5.69$\pm$0.01 & 5.82$\pm$0.01 &  \cite{2008AJ....136..735L} \\
- & 5.847$\pm$0.011 & \cite{2006yCat.2168....0M} \\
- & 5.846$\pm$0.002 &\cite{1998AAS..129..431H}\\
- & 5.844$\pm$0.017 & \cite{1995yCat.5086....0G}\\
5.740$\pm$0.004 & 5.850$\pm$0.003 & \cite{2001yCat.5109....0M} \\
5.687$\pm$0.014 & 5.818$\pm$0.009 & \cite{2006ApJ...638.1004A} \\
- & 5.845$\pm$0.006 & \cite{1992ApJS...82..117S} \\
\hline
\end{tabular}
%
\end{table}
We will calculate the extinction $\mathrm{A_{V}}$, effective temperatures T and bolometric corrections B.C. (as well as masses and ages) for B7V and B8V in the following. With the BVJHK colours and effective temperatures listed in \citet{1998A&A...333..231B} we obtain $\mathrm{T_{eff}=13000^{+1000}_{-1100}\,K}$ and $\mathrm{BC_{V}=-0.85\pm0.20\,mag}$ leading to $\mathrm{A_{V}=0.05 \pm 0.04\,mag}$ for a B7V star (using the absorption models by \citealt{1989ApJ...345..245C,1985ApJ...288..618R,1979ARA&A..17...73S}). For a B8V star we obtain $\mathrm{T_{eff}=11900^{+1100}_{-1400}\,K}$, $\mathrm{BC_{V}=-0.66\pm0.24\,mag}$ and $\mathrm{A_{V}=0.00\pm 0.04\,mag}$.\\
With magnitudes, $\mathrm{A_{V}}$ values, distances and bolometric corrections we obtain $\mathrm{L_{bol}=212\pm 65\,L_{\sun}}$ and $\mathrm{L_{bol}=170 \pm 67\,L_{\sun}}$ for B7V and B8V, respectively. With effective temperatures and luminosities, we now can estimate age and mass of HR\,3672 using the models for stellar evolution from \citet{1994A&AS..106..275B,2004yCat..34240919C} and \citet{1992A&AS...96..269S}. To account for possible uncertainties in the determination of the spectral type and different model temperatures we allow a relative error of one spectral type for the effective temperatures and search for the closest isochrones and mass-tracks in the model (within the error box from temperature and luminosity) and interpolate linearly between neighbouring data points (note, the model from \citealt{1992A&AS...96..269S} has 1\,100 data points, whereas the model from \citealt{2004yCat..34240919C} has 61\,000 data points).

\begin{figure}
 \includegraphics[width=84mm]{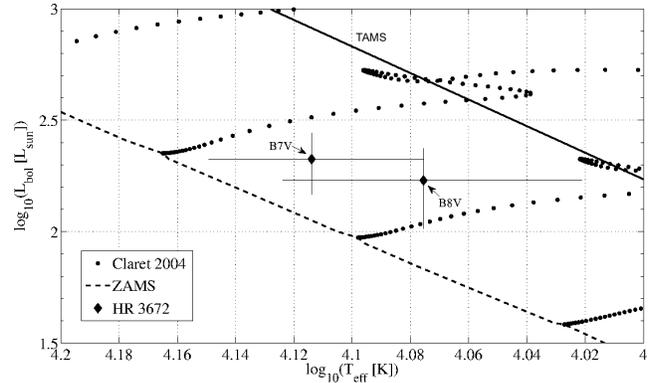}
\caption{Mass tracks in the stellar evolution models of \citet{2004yCat..34240919C}. The error-bars for the luminosity are calculated from the errors of the magnitude, parallax, extinction and bolometric correction and the error for the temperatures is assumed to be one spectral type. We also show the zero age main sequence (ZAMS, dotted line) and the termination age main sequence (TAMS, full line).}
\label{isochr_mass}
\end{figure}


First, we calculate the masses and ages for a B7V star. For solar metallicity we obtain $\mathrm{3.75\,M_{\sun}}$, $\mathrm{3.98\,M_{\sun}}$ and $\mathrm{4.00\,M_{\sun}}$ (for the models from \citealt{1994A&AS..106..275B,2004yCat..34240919C} and \citealt{1992A&AS...96..269S}, respectively), i.e. with median and standard deviation from all models we obtain $\mathrm{3.91\pm0.14\,M_{\sun}}$. With the same models we obtain the ages $\mathrm{79\,Myr}$, $\mathrm{27\,Myr}$ and $\mathrm{59\,Myr}$, and a median age of $\mathrm{59\pm37\,Myr}$, respectively. For a B8V star we get masses of $\mathrm{3.45\,M_{\sun}}$, $\mathrm{3.16\,M_{\sun}}$ and $\mathrm{3.00\,M_{\sun}}$ (with same order of the models), hence $\mathrm{3.20 \pm 0.23\,M_{\sun}}$ and ages of 158\,Myr, 225\,Myr and 223\,Myr (for the models from \citealt{1994A&AS..106..275B,2004yCat..34240919C} and \citealt{1992A&AS...96..269S}, respectively), hence $\mathrm{223\pm38\,Myr}$. These values for luminosity, temperature and mass are fully consistent with $\mathrm{log\,g=4}$ for a B\,7-8 main sequence star. In both cases, HR\,3672 lies well on the main sequence (see Fig. \ref{isochr_mass}), between ZAMS and TAMS. 

\begin{figure}
 \includegraphics[width=84mm]{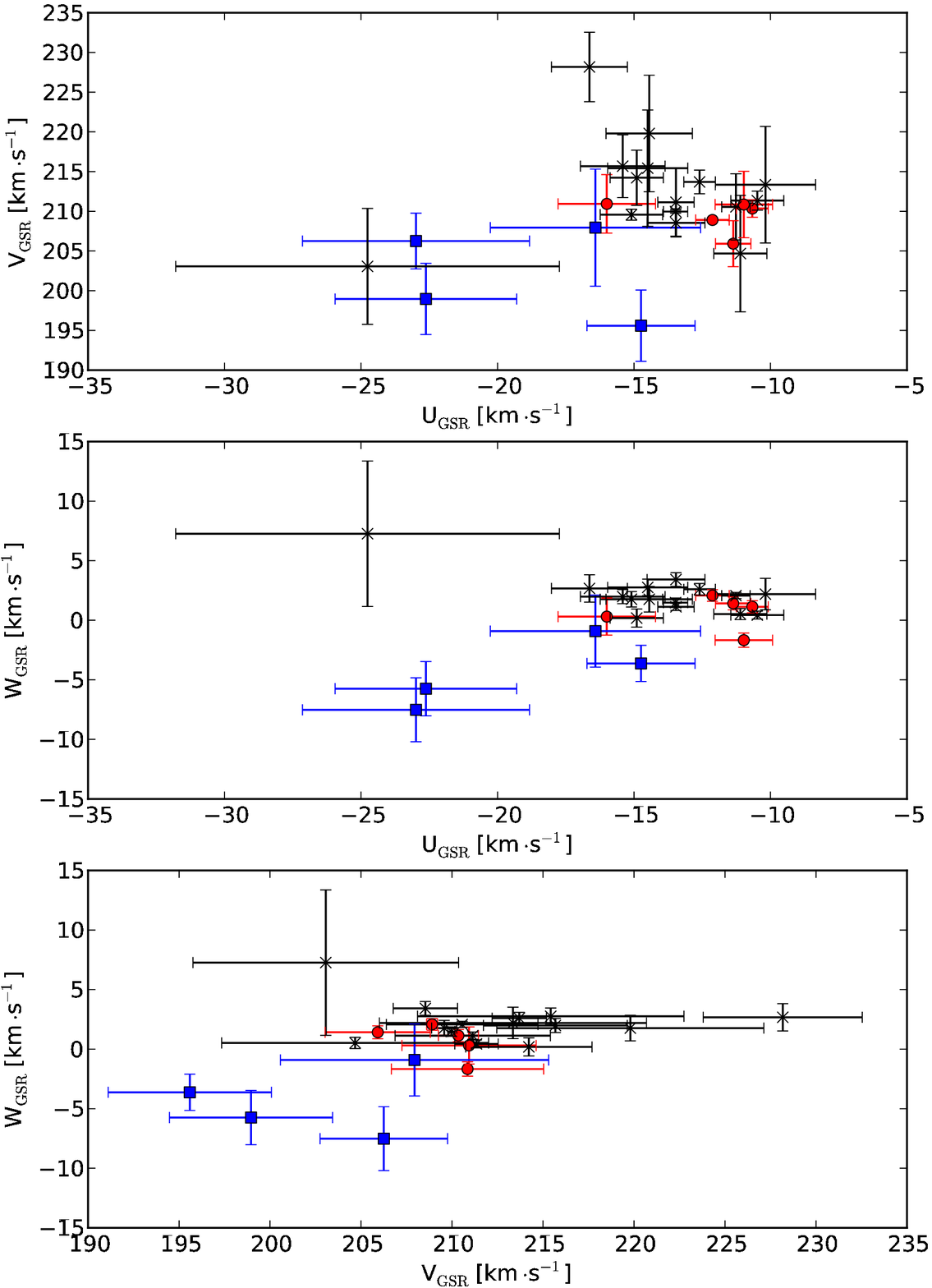}
\caption{U, V, and W heliocentric space motions (right-handed system) with errors for the members of Platais\,9 (circle), Trumpler\,10 (square) and IC\,2391 (asterisk) calculated from data according to \citet{2012yCat.5137....0A} (for details, see Tab.\,\ref{tabUVW}). The space motions are corrected for solar motion with respect to the local standard of rest (LSR) \citep{1998MNRAS.298..387D} and galactic rotation of the Sun \citep{1986MNRAS.221.1023K}. According to \citet{1998AJ....116.2423P}, HR\,3672 is a member of the possible HIP\,45189 cluster. The results from our calculations show that the members of the Platais\,9 cluster, including HR\,3672, do cluster in UVW and space. Similar UVW results can be obtained for Trumpler\,10 and IC\,2391, but locations in space are different.}
\label{uvw_vel}
\end{figure}

HR\,3672 was identified as probable member of the possible open cluster HIP\,45189 by \citet{1998AJ....116.2423P} and is located in or near the Vela OB association. Using the radial velocity measurements, proper motions and trigonometric parallaxes we can check, whether the Platais 9 cluster according to \citet{1998AJ....116.2423P} with previous Hipparcos data \citep{1997ESASP1200.....P} is still a cluster with the revised Hipparcos data \citep{2007A&A...474..653V} and radial velocties. We have calculated the U, V and W heliocentric velocity components in the directions of the Galactic center, Galactic rotation, and north Galactic pole, respectively, with the formulation developed by \citet{1987AJ.....93..864J} for the members of the Platais 9, Trumpler 10 and IC\,2391 cluster, as shown in Fig. \ref{uvw_vel}. Note that the right-handed system is used and we applied corrections for solar motion with respect to the local standard of rest (LSR) \citep{1998MNRAS.298..387D} and galactic rotation of the sun ($\mathrm{V_{gal,\sun}=220\,km/s}$, see \citet{1986MNRAS.221.1023K}).\\
Fig.\,\ref{uvw_vel} shows that the 5 stars of the cluster Platais 9 (including our star HR\,3672), for which both proper motions from Hipparcos and radial velocity \citep{2006AstL...32..759G} are known, do cluster in all three velocities (and in location in space anyway).
The stars of the cluster Trumpler\,10 and IC\,2391 also shown in Fig.\,\ref{uvw_vel} also do form a cluster in all 3 velocities, which are similar to Platais 9, but the locations in space are different. With this results from our calculations we can confirm the conclusions from \citet{1998AJ....116.2423P}, that HR\,3672 is a member of the cluster called Platais\,9.\\
Using colour-magnitude-diagrams (CMD) and theoretical isochrones from Yale Rotating Evolution Code \citep{2008ApSS.316...31D} \citet{1998AJ....116.2423P} found an age for the cluster HIP\,45189 and HR\,3672\,A of $\rmn{\sim\,100\,Myr}$, but applied no correction for reddening, i.e. the objects may be shifted to blue colours and, hence, lower ages.\\
HR\,3672 has a non-detection in the ROSAT All-Sky Survey with an upper limit to the X-ray luminosity being $\rmn{3.006\,\mu W}$, i.e. $\rmn{log(L_x/L_{bol})\,<\,-6.45}$ \citep{1996A&AS..118..481B}. This ratio gives only a very high age upper limit of several Gyr \citep{2008ApJ...687.1264M}. \textbf{This consideration holds, if the X-ray emission would come from the primary. If the emission would come only from the secondary, and we assume the maximum X-Ray emission for M-type stars (see below) with $\rmn{log(L_x/L_{bol})\,\sim\,-3}$ \citep{1993ApJ...410..387F}, then the upper age limit would be roughly $\rmn{\approx\,70\,Myr}$ according to X-ray - Age correlation by \citet{2005ApJS..160..390P} for $\rmn{0.2-0.4}$ solar mass stars. It also turns out, that even at this young age the X-ray luminosity of HR\,3672\,B is too low for detection.}

\subsection{Mass of HR\,3672\,B}
\label{mass_B}
Since the only observable for the determination of the mass of HR\,3672\,B is the magnitude difference in the K-band, we calculate the absolute K-band magnitude of HR\,3672\,A for a B7V and B8V star, respectively, taking the absorption in the K-band $\mathrm{A_{K}}$ into account. With the absorption models by \citet{1989ApJ...345..245C}; \citet{1985ApJ...288..618R} and \citet{1979ARA&A..17...73S} and the values for $\mathrm{A_{V}}$ (see above) we obtain $\mathrm{A_{K}=0.006 \pm 0.005\,mag}$ for a B7V star and $\mathrm{A_{K}=0.000 \pm 0.005\,mag}$ for a B8V star, respectively. Using the apparent K-band magnitude $\mathrm{K_{s}=6.134\pm0.027\,mag}$ \citep{2003yCat.2246....0C} and distance we obtain an absolute K-band magnitude of $\mathrm{M_{K}=0.10\pm0.24\,mag}$ for HR\,3672\,A\,+\,B together (unresolved).\\
With $\mathrm{\Delta K_{s}=6.28\pm0.13\,mag}$ between HR\,3672\,B and A, and distance, we obtain $\mathrm{M_{K,A}=0.099\pm0.236\,mag}$ for HR\,3672\,A, and $\mathrm{M_{K,B}=6.38\pm0.48\,mag}$ for the B-component in K-band. At $\mathrm{59\pm37\,Myr}$, for a B7V star, we derive a mass of $\mathrm{0.30\pm0.1\,M_{\sun}}$ and $\mathrm{0.3\pm0.1\,M_{\sun}}$ from evolutionary models by \citet{1998A&A...337..403B} and \citet{2000A&A...358..593S}. In case of a B8V star we obtain an age of $\mathrm{223\pm38\,Myr}$ and masses of $\mathrm{0.40\pm0.05\,M_{\sun}}$ and $\mathrm{0.4\pm0.1\,M_{\sun}}$ from \citet{1998A&A...337..403B} and \citet{2000A&A...358..593S}, respectively. Using empirical \textbf{relations} by \citet{2000A&A...364..217D} and the absolute magnitude in K-band, we derive a mass of $\mathrm{0.38\pm0.05\,M_{\sun}}$. Hence, HR\,3672\,B would be a low-mass stellar object of spectral type M.

\subsection{Limits on further companions}
\label{det_limit}

\begin{figure}
 \includegraphics[angle=0,width=84mm]{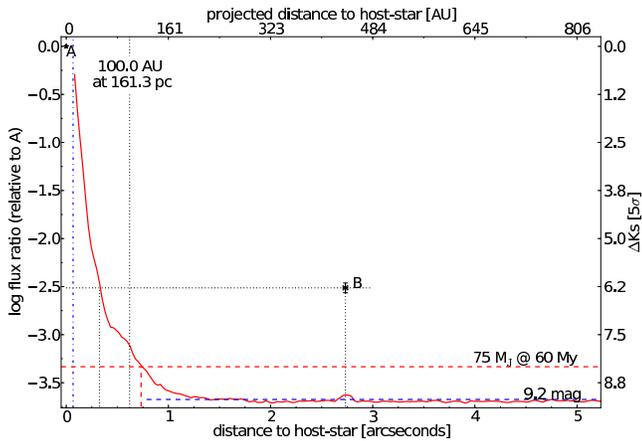}
  \caption{Dynamic range of HR\,3672 for the co-added images from 2004-12-16 and 2008-02-15 in Ks-band. We plot the log flux ratio between background and HR\,3672\,A as function of the separation in arc-seconds. The dashed-dotted line (blue) indicates the defraction limit of the VLT/UT4 ($D = 8.2\,\rmn{m}$) in the Ks-band ($\lambda = 2.18\,\rmn{\mu m}$). The full (red) line shows the measured $5\,\sigma$ noise-level. Objects below this line cannot be detected. The background-level remains constant at $\rmn{\Delta K_{s}= 9.2\,mag}$ for separations outside of $1.5\,\arcsec$ ($\mathrm{0.05\,M_{\sun}}$ at $\mathrm{60\,Myr}$). At $100\,\rmn{AU}$ ($0.6\,\arcsec$ at $161.3\,\rmn{pc}$), we could detect companions with $7.7\,\rmn{mag}$ difference ($\mathrm{0.11\,M_{\sun}}$ at $\mathrm{60\,Myr}$); other companions as faint as HR\,3672\,B could be detected for $\geq0.3\,\arcsec$ (or $48\,\rmn{AU}$). Furthermore we show the brown-dwarf limit ($75\,M_{\rmn{J}}$) for $60\,\rmn{Myr}$ as dashed line (red). Hence, sub-stellar objects were detectable outside of $0.7\,\rmn{\arcsec}$ (or $\sim112\,\rmn{AU}$).}
  \label{dyn_range}
\end{figure}

For the co-added image (VLT NACO 2004+2008), no additional companion candidates were detected within $13\,\rmn{\arcsec}$. Assuming an age of $\mathrm{\sim60\,Myr}$ (see Sec.\,\ref{mass_age_A}), sub-stellar objects would be detectable outside of $\mathrm{0.7\,\arcsec}$ (or $\sim112\,\rmn{AU}$). For an age of $\rmn{1\,Myr}$, sub-stellar objects would be detectable outside of $\mathrm{0.2\,\arcsec}$. Hence, HR\,3672\,B would be a sub-stellar object.

The determination of the dynamic range is done by measurement of the $5\,\sigma$ level above the background noise in the co-added image from both epochs. We compared the measured background flux from groups of $\mathrm{3\times3}$ pixels with the flux of the central star HR\,3672\,A. In Fig.\,\ref{dyn_range} we plot the flux ratio between background and HR\,3672\,A for the co-added images.

\section{conclusions}
\label{conclusion}
From the archival data of HR\,3672/HIP\,45314 obtained with VLT NACO with $4\,\rmn{yrs}$ epoch difference, we could reject ($>3\,\sigma$) the background hypothesis, that HR\,3672\,B is unrelated to HR\,3672\,A. Hence, HR\,3672\,A and B form a common proper motion pair. The small change in separation and the irregularity of the position angle may indicate orbital motion. No additional companions were detected within $\mathrm{13\arcsec}$.\\
Using unpublished IUE UV-spectra, we could narrow the spectral type of HR\,3672\,A down to a B7-8 main-sequence star. With evolutionary models from \citet{1994A&AS..106..275B,2004yCat..34240919C} and \citet{1992A&AS...96..269S}, respectively, we derive a mass range of 3 to 4 $\rmn{\,M_{\sun}}$ for HR\,3672\,A. From models we obtain ages of $\rmn{\sim\,60\,Myr}$ for a B7V star, and $\rmn{\sim\,220\,Myr}$ for a B8V star, respectively.\\
By comparison of U, V, W heliocentric space motion with surrounding cluster we can confirm HR\,3672 to be a member of the cluster Platais 9 \citep{1998AJ....116.2423P} with an age of about 100$\mathrm{\,Myr}$.\\
The magnitude difference between HR\,3672\,A and B is $\rmn{\Delta K_{s}=6.28\pm0.13\,mag}$, and with $\rmn{K_{s}=6.134\pm 0.027\,mag}$ (2MASS) for HR\,3672\,A+B, we get $\rmn{K_{s}=12.42\pm0.28\,mag}$ for HR\,3672\,B. Given the magnitude difference and age of HR\,3672\,A of 60 to 220\,Myr we derive a mass of 0.15 to 0.5$\rmn{\,M_{\sun}}$ from evolutionary models (e.g. \citet{1998A&A...337..403B}, or \citet{2000A&A...358..593S}) and empirical mass-luminosity relations by \citet{2000A&A...364..217D}. Hence, HR\,3672\,B is a low mass star. The high uncertainties in mass of B can be decreased by taking spectra of HR\,3672\,B to get the physical parameters with small uncertainties.

\section*{Acknowledgements}
RN , CA and TOBS wish to acknowledge Deutsche Forschungsgemeinschaft (DFG) for grant NE 515/35-1. JS wish to thank SFB for grant. This research has made use of the SIMBAD database and the VizieR catalogue access tool, operated at CDS, Strasbourg, France and archival data from the ESO, 2MASS and INES (ESA IUE Project). 


\newpage

\appendix

\section{Plots}
For completeness we show the results of self-calibration for pixel scale and detector orientation for all epochs with same remarks as Fig.\,\ref{ps_calib_2004_img}. 


\begin{figure}
 \includegraphics[width=84mm]{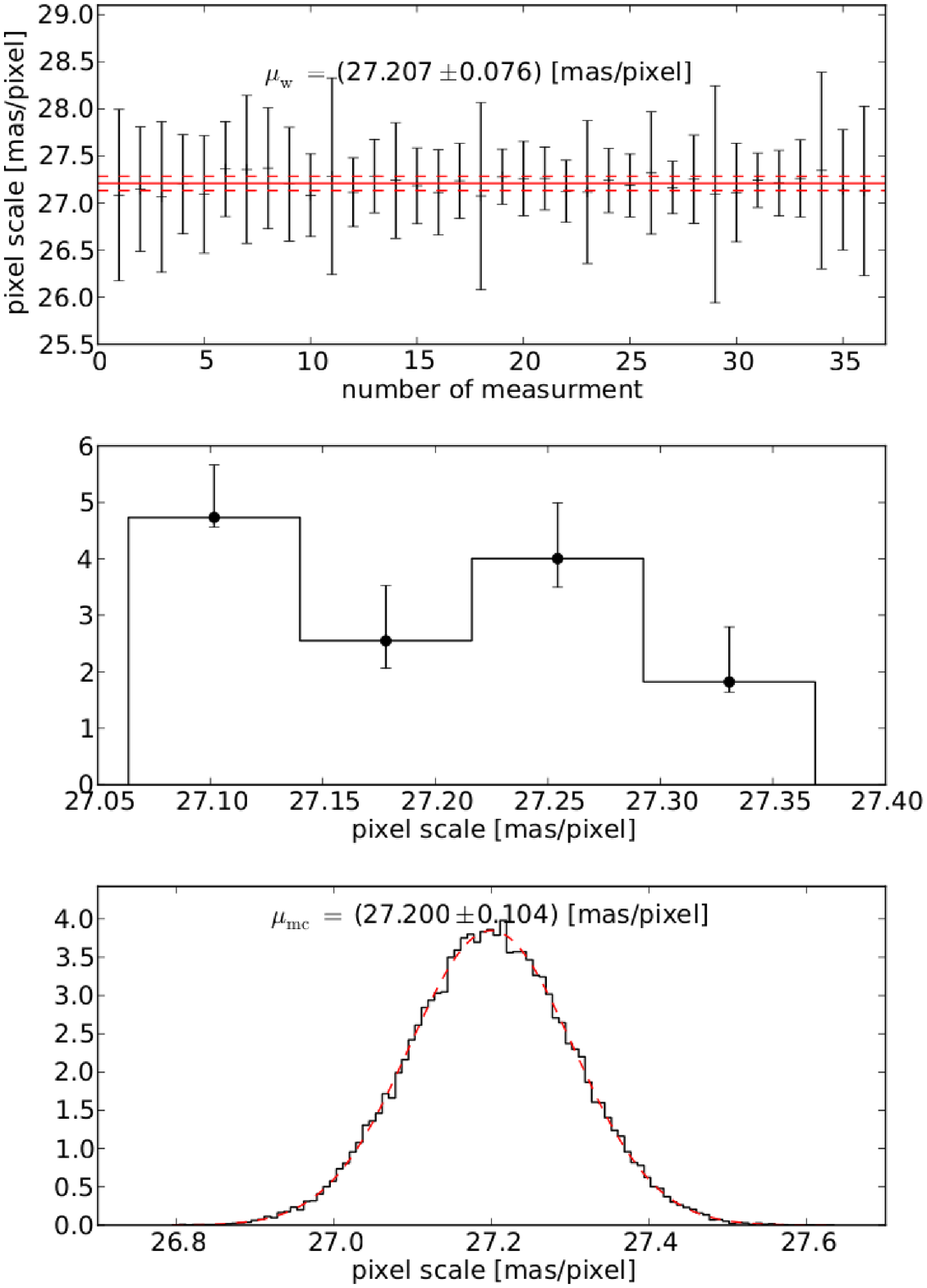}
\caption{self-calibration result; pixel scale for 2008/02/15}
\end{figure}

\begin{figure}
 \includegraphics[width=84mm]{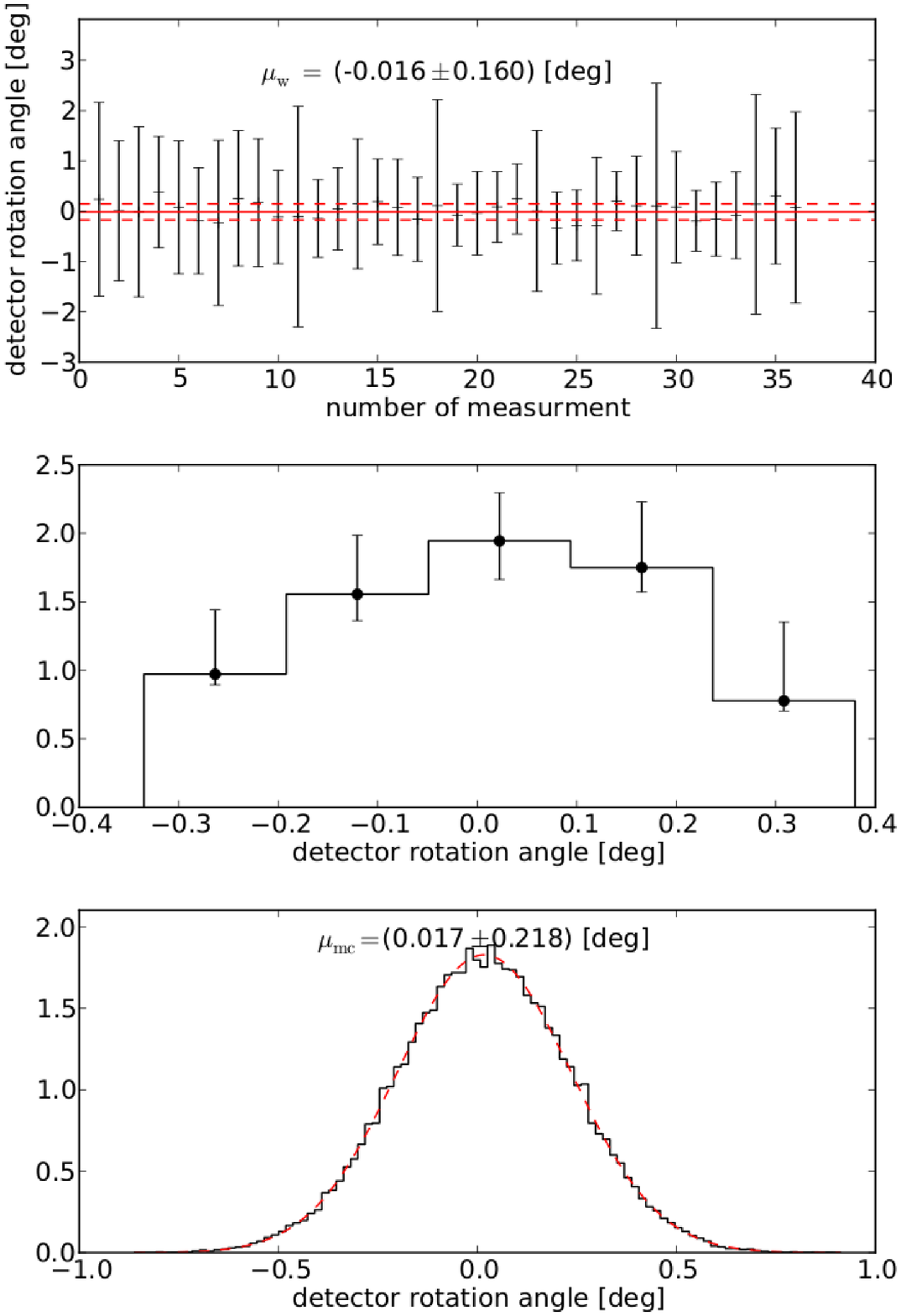}
\caption{self-calibration result; detector orientation for 2008/02/15}
\end{figure}

\begin{figure}
 \includegraphics[width=84mm]{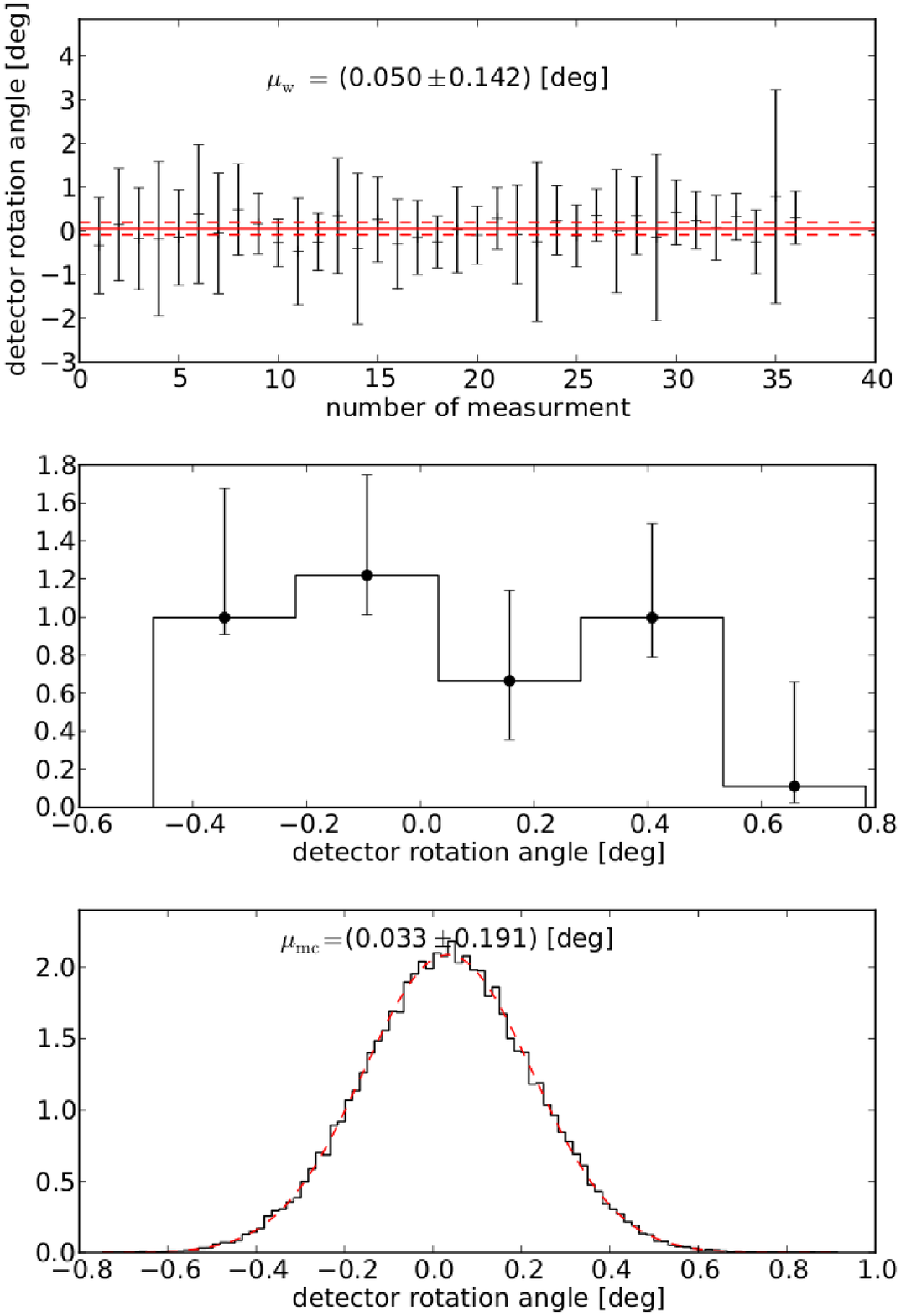}
\caption{self-calibration result; detector orientation for 2004/12/16}
\end{figure}

\newpage
\section{Tables}

\begin{table*}
 \begin{minipage}{168mm}
\caption{Kinematic Parameter for Members of selected Cluster for UVW space velocity}
\label{tabUVW}
\begin{tabular}{@{}lcccccccc}
 \hline
Group & HIP & R.A.~(J2000.0) & Dec.~(J2000.0) & $\rmn{\pi}$ & $\rmn{\mu_{\alpha}}$ & $\rmn{\mu_{\delta}}$  & RV  & Reference  \\
 &   & deg & deg & mas & mas~$yr^{-1}$ & mas~$yr^{-1}$ & km~$s^{-1}$    & for RV  \\
 \hline
Platais 9 & 44816 & 136.999 & -43.433 & 5.99$\pm$0.11 & -23.4$\pm$0.70 & 14.4$\pm$0.7 & 17.6$\pm$0.3     &  1  \\
 & 45189 & 138.127 & -43.613 & 4.71$\pm$0.46 & -22.72$\pm$0.48 & 12.87$\pm$0.38 & 15.4$\pm$3.7          &  2 \\
 & 45270 & 138.394 & -47.338 & 6.40$\pm$0.25 & -25.23$\pm$0.25 & 14.24$\pm$0.23 & 15.0$\pm$1.1             &  1  \\
 & 45314 & 138.534 & -44.146 & 6.20$\pm$0.27 & -24.90$\pm$0.26 & 13.13$\pm$0.19 & 20.0$\pm$2.9              &  1  \\
 & 45344 & 138.602 & -43.228 & 5.33$\pm$0.26 & -23.6$\pm$0.9 & 8.7$\pm$0.9 & 15.0$\pm$4.2                &   1 \\
 & 45395 & 138.766 & -48.504 & 5.00$\pm$0.63 & -22.43$\pm$0.56 & 13.17$\pm$0.51                     &    \\
 & 45820 & 140.138 & -44.283 & 4.74$\pm$0.56 & -24.2$\pm$1.0 & 13.8$\pm$1.1                         &    \\
 & 46024 & 140.796 & -44.965 & 5.54$\pm$0.48 & -25.36$\pm$0.43 & 13.58$\pm$0.36                  &    \\
 \hline
Trumpler 10 & 42939 & 131.293 & -41.427 & 2.22$\pm$0.86 & -12.89$\pm$0.59 & 7.17$\pm$0.66  &    \\    
 & 43055 & 131.574 & -42.759 & 2.54$\pm$0.75 & -11.99$\pm$0.57 & 7.00$\pm$0.58             &    \\
 & 43087 & 131.643 & -42.791 & 3.09$\pm$0.95 & -12.34$\pm$0.67 & 8.20$\pm$0.71        &    \\
 & 43128 & 131.772 & -45.075 & 2.81$\pm$0.33 & -12.85$\pm$0.30 & 6.34$\pm$0.25 & 32.0$\pm$4.5             &  1  \\
 & 43182 & 131.95 & -42.273 & 2.82$\pm$0.49 & -12.86$\pm$0.37 & 6.13$\pm$0.40       &    \\
 & 43209 & 132.037 & -42.463 & 2.07$\pm$0.30 & -12.50$\pm$0.27 & 6.22$\pm$0.24 & 30.0$\pm$4.5              &  1  \\
 & 43285 & 132.264 & -43.761 & 2.63$\pm$0.53 & -12.04$\pm$0.50 & 7.68$\pm$0.41 & 20.0$\pm$7.4             &  2  \\
 & 43520 & 132.959 & -44.151 & 1.94$\pm$0.33 & -12.75$\pm$0.34 & 5.93$\pm$0.25 & 22.0$\pm$3.5            &  1  \\
 \hline
IC 2391 & 41644 & 127.374 & -54.212 & 4.47$\pm$0.28 & -16.6$\pm$1.0 & 17.7$\pm$1.0 & 10.1$\pm$4.0            &  3  \\
 & 42121 & 128.792 & -54.206 & 4.12$\pm$0.33 & -14.72$\pm$0.32 & 15.77$\pm$0.34 & 10.0$\pm$7.4           &  2  \\
 & 42216 & 129.098 & -53.037 & 4.10$\pm$1.19 & -15.8$\pm$1.3 & 25.7$\pm$1.3 & 22.0$\pm$7.4                &  2  \\
 & 42274 & 129.294 & -53.259 & 6.46$\pm$0.57 & -24.96$\pm$0.55 & 23.0$\pm$0.5 & 6.0$\pm$7.4                &  2  \\
 & 42374 & 129.6 & -53.722 & 7.06$\pm$0.43 & -23.72$\pm$0.46 & 22.00$\pm$0.42 & 21.0$\pm$7.4                &  2  \\
 & 42400 & 129.687 & -53.09 & 6.23$\pm$0.33 & -25.53$\pm$0.36 & 21.84$\pm$0.3 & 11.7$\pm$3.5           &  1  \\
 & 42450 & 129.819 & -52.314 & 5.55$\pm$0.69 & -24.69$\pm$0.65 & 22.41$\pm$0.57 & 12$\pm$999           &  4  \\
 & 42459 & 129.849 & -53.44 & 7.33$\pm$0.24 & -23.90$\pm$0.24 & 23.46$\pm$0.22 & 15.0$\pm$4.2            &   1 \\
 & 42504 & 129.99 & -53.055 & 7.37$\pm$0.34 & -25.5$\pm$1.2 & 20.8$\pm$1.0 & 14.5$\pm$1.2              &    5\\
 & 42535 & 130.073 & -53.015 & 6.79$\pm$0.26 & -25.30$\pm$0.27 & 23.14$\pm$0.24 & 14.6$\pm$4.3        &    1\\
 & 42536 & 130.073 & -52.922 & 6.61$\pm$0.35 & -25.1$\pm$0.9 & 25.0$\pm$0.9 & 16.1$\pm$0.7            &  1  \\
 & 42702 & 130.542 & -52.968 & 6.06$\pm$0.43 & -25.60$\pm$0.46 & 23.23$\pm$0.42 & -2.6$\pm$4.4        &  1  \\
 & 42714 & 130.576 & -53.902 & 7.81$\pm$0.94 & -25.05$\pm$0.95 & 23.22$\pm$0.87 & 12.0$\pm$7.4         &  2  \\
 & 42715 & 130.579 & -53.1 & 6.53$\pm$0.26 & -21.8$\pm$1.2 & 24.3$\pm$1.1 & 16.8$\pm$1.78           &   2 \\
 & 42726 & 130.606 & -53.114 & 6.73$\pm$0.17 & -24.84$\pm$0.18 & 23.23$\pm$0.18 & 15.6$\pm$0.2      &   5 \\
 & 42809 & 130.842 & -52.209 & 2.83$\pm$0.68 & -23.92$\pm$0.60 & 22.00$\pm$0.55    &    \\
 & 42823 & 130.887 & -52.004 & 6.90$\pm$0.59 & -27.57$\pm$0.52 & 26.32$\pm$0.50       &    \\
 & 43071 & 131.597 & -52.844 & 5.25$\pm$0.51 & -21.5$\pm$1.6 & 22.5$\pm$1.4       &    \\
 & 43195 & 132.001 & -52.85 & 7.05$\pm$0.24 & -25.10$\pm$0.25 & 23.31$\pm$0.25 & 11.6$\pm$1.5        &   1 \\
 & 43433 & 132.696 & -54.113 & 5.75$\pm$0.58 & -23.1$\pm$2.2 & 19.6$\pm$1.8         &    \\

\hline
\end{tabular}

\medskip
{\bf Note.} Kinematic data for Cluster member stars according to \cite{2012yCat.5137....0A} with radial velocities from 1: \cite{2006AstL...32..759G}, 2: \cite{2007AN....328..889K}, 3: \cite{1996A&AS..118..231L}, 4: \cite{1993BICDS..43....5T} and 5: \cite{2004A&A...424..727P}.
 \end{minipage}
\end{table*}

%


\label{lastpage}
\end{document}